\documentclass[final,5p,times]{elsarticle}

\usepackage{graphicx}
\usepackage{amssymb}

\usepackage{lineno}

\usepackage{comment}
\usepackage{amsmath}
\usepackage{lipsum}
\usepackage{graphics}
\usepackage[dvipsnames]{xcolor}
\usepackage{ulem}
\usepackage{newtxmath}
\usepackage{hyperref}

\newcommand{\beq}{\begin{equation}}
\newcommand{\eeq}{\end{equation}}

\def\vec   #1{\mbox{\boldmath $#1$}{}}
\def\scas  #1{\mbox{{\scriptsize{${\rm{#1}}$}}}{}}

\DeclareMathOperator*{\argminA}{arg\,min}
\DeclareMathOperator*{\argmaxA}{arg\,max}

\journal{ArXiv}

\begin{document}

\begin{frontmatter}


\title{Improving Reconstructive Surgery Design using Gaussian Process Surrogates \\ to Capture Material Behavior Uncertainty}



\author{Casey Stowers $^1$, Taeksang Lee$^1$, Ilias Bilionis$^1$, Arun K Gosain$^{3}$, Adrian Buganza Tepole$^{1,3}$}

\address{$^1$School of Mechanical Engineering, Purdue University, West Lafayette, IN, USA\\ $^2$Lurie Children Hospital, Northwestern University, Chicago, IL, USA\\$^3$Weldon School of Biomedical Engineering, Purdue University, West Lafayette, IN, USA}

\begin{abstract}
To produce functional, aesthetically natural results, reconstructive surgeries must be planned to minimize stress because excessive loads near wounds have been shown to produce pathological scarring and other complications \cite{gurtner2011improving}. Presently, stress cannot easily be measured in the operating room. Consequently, surgeons rely on intuition and experience \cite{paul2016new,buchanan2016evidence}. Predictive computational tools are ideal candidates for surgery planning. Finite element (FE) simulations have shown promise in predicting stress fields on large skin patches and complex cases, helping to identify potential regions of complication. Unfortunately, these simulations are computationally expensive and deterministic \cite{lee2018b}. However, running a few, well selected FE simulations allows us to create Gaussian process (GP) surrogate models of local cutaneous flaps that are computationally efficient and able to predict stress and strain for arbitrary material parameters. Here, we create GP surrogates for the advancement, rotation, and transposition flaps. We then use the predictive capability of these surrogates to perform a global sensitivity analysis, ultimately showing that fiber direction has the most significant impact on strain field variations. We then perform an optimization to determine the optimal fiber direction for each flap for three different objectives driven by clinical guidelines \cite{leedy2005reconstruction,rohrer2005transposition} . While material properties are not controlled by the surgeon and are actually a source of uncertainty, the surgeon can in fact control the orientation of the flap with respect to the skin's relaxed tension lines, which are associated with the underlying fiber orientation \cite{borges1984relaxed}. Therefore, fiber direction is the only material parameter that can be optimized clinically. The optimization task relies on the efficiency of the GP surrogates to calculate the expected cost of different strategies when the uncertainty of other material parameters is included. We propose optimal flap orientations for the three cost functions and that can help in reducing stress resulting from the surgery and ultimately reduce complications associated with excessive mechanical loading near wounds.

\end{abstract}

\begin{keyword}
Nonlinear finite elements \sep Local flaps \sep Soft tissue mechanics \sep Machine learning \sep Skin biomechanics 

\end{keyword}

\end{frontmatter}

\section*{Introduction}\label{motiv}
Reconstructive surgery requires balancing long term tissue functionality while producing aesthetically natural results \cite{gosain2009,buchanan2016evidence,aarabi2007hypertrophic}. Complications such as wound dehiscence, pathological scarring, and skin necrosis are partially caused by excess stress, clinically referred to as tension, especially along suture lines \cite{gurtner2011improving,paterno2011akt,wong2011mechanical,wong2012focal,rustad2013role}. However, measuring stress in the operating room is not practically possible to date with the exception of research studies \cite{raposio1998undermining,paul2016new}. Thus, surgeons rely on intuition built from experience and training to estimate skin tension and plan the surgery \cite{leedy2005reconstruction,norman2009nonmelanoma,maciel2013local,buchanan2016evidence}. This approach is not quantitative, making it difficult to train residents on objective metrics. In addition, the skin tension is estimated at a point in the process where changes to the surgical plan are no longer feasible - for example, once surgical excision of a skin lesion has already occurred. In view of these limitations, high-fidelity computational models of tissue mechanics can be used to recreate virtual surgery scenarios and estimate the stress distribution from a given surgical plan \cite{lee2018b,buganza14a}. However, a longstanding limitation of computational models is the difficulty to incorporate inherent variability and uncertainty of skin mechanical behavior between individuals \cite{luebberding2014,lee2019}. Another challenge is that high-fidelity models are too computationally expensive for routine clinical use \cite{rajabi2015rhombic,mitchell2016real}. Here, we seek to overcome these problems in the context of local cutaneous flaps by creating Gaussian Process (GP) surrogate models from detailed finite element (FE) models. The surrogate models are computationally inexpensive, yet they are accurate over a wide range of material parameters, including anisotropy. We show that these GP surrogates, being computationally affordable, can be leveraged to easily solve flap optimization tasks requiring a large number of function evaluations.  

\begin{figure*}[!htb]
\begin{center}
\includegraphics[width=0.7\textwidth]{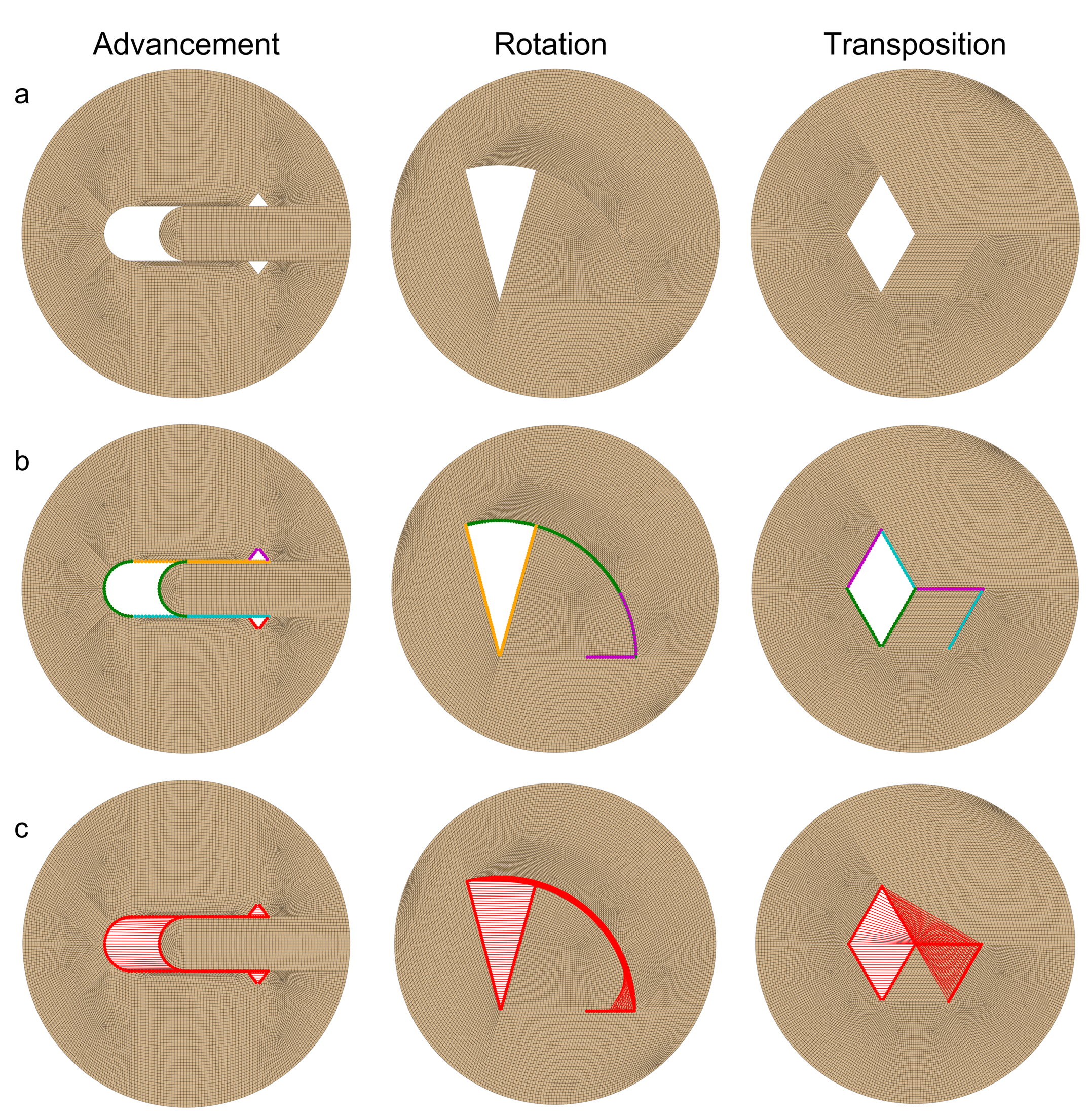}
\caption{The three most common flap designs are advancement, rotation and transposition flaps. FE models of the flaps are generated semi-automatically. a) Base flap design generated manually in Abaqus. b) Automatic identification of the edges. Matching colored edges in each design are brought together by sutures to close the skin. Adjacent skin regions to the edges are also identified and used to impose essential boundary conditions. In this case, only the outer perimeter is fixed. c) Suturing pattern imposed as linear constraints between pairs of nodes are applied gradually to bring the flap together. In this case, the sutures are mapped across the paired edges such that they have a uniform spacing.}
\label{fig01}
\end{center}
\end{figure*}

Local flaps are commonly used to repair cutaneous lesions such as skin cancer or burn lesions \cite{norman2009nonmelanoma,madan2010non,gumucs2013management}. Local flaps have several aesthetic advantages over other reconstructive techniques, including having the same color, hair bearing properties, and blood supply as the skin surrounding the lesion. Local flap designs can be classified based on their geometry, i.e., the  pattern of the incision and the pairs of points along the edges of the flap that are brought together by sutures \cite{patel2011concepts, paul2016clinical}. While new flap designs are still being proposed \cite{paul2016new,arpaci2017omega}, the most common are currently the advancement, rotation, and transposition flaps \cite{buchanan2016evidence,patel2011concepts} (see Figure \ref{fig01}). Here, we start from these three flap designs and create detailed FE models to predict the resulting strain contours over the flap and surrounding skin. Strain is selected as the quantity of interest in this study because, unlike stress, strains can be measured non-invasively even in the operating room with the use of 3D photography or multi-view stereo (MVS) \cite{buganza15,lee2018multi}. Moreover, strain measurements have been linked to compression of the flap in the thickness direction and collapse of the microvasculature leading to ischemia and subsequent complications \cite{barnhill1984study}. 

Previous examples of FE simulations of reconstructive surgery have improved our understanding of basic features of stress and strain profiles for common flaps. Additionally, they have already gained recognition as a promising tool for prediction of potential healing complications in personalized cases \cite{buganza14a,remache2015numerical,lee2018multi}. However, as mentioned above, FE models are computationally expensive and deterministic in nature, while flap parameters and skin material properties in clinical settings entail unavoidable uncertainties that cannot be captured with a single FE evaluation \cite{krueger2011,annaidh2012b,tonge2013}. Instead, many FE simulations would be required to propagate uncertainties through the model, which is not feasible in routine clinical settings. Optimization of flap design using FE simulations is also challenging on a budget  \cite{rajabi2015rhombic}. Thus, FE simulations are not ready for clinical use. Unfortunately, this means that current recommendations are described in qualitative terms such as \textit{feel} or \textit{manual estimation} of tension \cite{leedy2005reconstruction}.

Machine learning tools can be leveraged to reduce the computational burden of detailed computational simulations by learning inexpensive metamodels of the original high-fidelity models \cite{latorre2018,lejeune2020mechanical}. For example, stress and strain features as a function of different inputs can be learned using machine learning tools such as GPs \cite{lee2019,peirlinck2019using}. This and other machine learning approaches have been used in other fields of computational physics applied to medical problems, for example to replace computational fluid dynamics (CFD) models with deep neural networks \cite{perdikaris2016model}, or to capture cardiac electrophysiology \cite{costabal2019multiscale,costabal2019}. A recent perspective on the integration of machine learning methods and physics-based models covers more examples, and also highlights the current challenges and opportunities in this field \cite{peng2020multiscale}.

Here, we first obtain a reduced order representation of the strain fields through principal component analysis (PCA). Then, the data consisting of the reduced output with its corresponding input (material parameters including anisotropy), is used to train a computationally efficient GP surrogate. The GP surrogates are trained on a few well-selected FE simulations, but predict stress and strain accurately for arbitrary material properties within a broad range. 

The work shown in this paper extends the previous work in this area through considering anisotropy, which is a key feature of soft tissues, including skin \cite{gasser2005,annaidh2012}. More importantly, in the clinical scenario, most of the material parameters describing skin's mechanical behavior are highly uncertain except for the anisotropy direction. Starting from the seminal work of Langer in the nineteenth century, the anatomy of skin anisotropy  has been well documented \cite{langer1978anatomy,langer1978anatomy2,reihsner1995two}. The relaxed skin tension lines, which are associated with the underlying fiber direction, are the most common anisotropy feature used surgical planning \cite{borges1984relaxed}. Our sensitivity analysis reveals that the anisotropy direction is the most important material parameter affecting the final strain distribution. Since the surgeon can actually control the orientation of the flap with respect to the relaxed skin tension lines, which we consider indicative of fiber direction, we optimize the flap orientation posing objective functions that reflect  clinical guidelines \cite{leedy2005reconstruction,rohrer2005transposition,2014GaSp}.

\section*{Methods}
\subsection*{Automating generation of FE models}
We create a base model of the incision geometry for each of the three flaps using Abaqus Standard (Simulia, Boston, USA). Unlike previous work \cite{lee2019}, we restrict our attention to a circular, two dimensional domain. The plane stress formulation is accurate for modeling skin because it is a thin membrane \cite{meador2020regional,kumaraswamy2017mechanical}, although more detailed models accounting for the multiple skin layers are also possible \cite{flynn2010simulating,groves2012quantifying}. In the cutaneous flaps shown here, the loading is expected to be within the skin plane and primarily under tension. Therefore, the dermis is expected to be the major contributor to the mechanical response of the tissue \cite{jor2011,benitez2017}, and the contributions of the other two skin layers, epidermis and hypodermis, are ignored. In previous work, setting up the incision geometry, boundary conditions and material parameters was done manually. This is a time consuming, inefficient process. Our long term goal is to bypass manual model creation. As a first step in this direction, we automate input file generation with a Python script which takes inputs of a few key points and the original Abaqus input file. The script automatically identifies the flap edges, generates the suturing scheme, and imposes boundary conditions taking into account the flap edges. Automating further the geometry of the incision pattern is a logical next step, but not explored in this paper. We focus on the material behavior uncertainty and consider a single suturing scheme and a single set of fixed boundary conditions. A wider set of simulations varying the flap design more broadly is ongoing work. For the simulations in this paper only the outer edge is fixed. The suturing scheme is illustrated in Figure \ref{fig01}. Sutures are imposed as linear constraints between pairs of nodes, which are gradually imposed to bring the corresponding edges together, analogous to the actual procedure \cite{buganza14a}. The script for input file generation is available with this submission. To aid in convergence, we allowed some of the simulations to add dissipation for stabilization of the solver. While this introduces an error in the prediction, we enforced this error to be very small since only a small number of simulations required this artificial dissipation and we kept the artificial energy as low as possible.

\subsection*{Constitutive model of skin}
The mechanical response of the dermis dictates the overall behavior of skin under tension \cite{limbert2017,lakhani2020directional}. The dermis is a collagenous tissue and can be considered nearly incompressible and hyperelastic \cite{jor2011}. Though it was originally developed to capture the mechanical behavior of arteries, the strain energy function proposed by Gasser-Ogden-Holzapfel (GOH) \cite{gasser2005} has been adapted to model skin and is used here \cite{annaidh2012,tonge2013}. The main features of this strain energy are its exponential behavior, and its microstructurally-inspired decomposition which naturally incorporates the influence of anisotropy due to collagen fiber networks. We briefly define the important kinematic quantities before introducing the strain energy. The deformation gradient $\vec{F}=\nabla_X {\bf \varphi}$ is the main kinematic object, capturing the local changes in geometry induced by the deformation map $\vec{x}={\bf \varphi}(\vec{X})$. Due to its nearly incompressible behavior, it is advantageous to split the deformation gradient into its isochoric and volumetric contributions 

\beq 
\label{eq1}
\vec{\bar{F}} = J^{-1/3} \vec{F}, \; \vec{F}^{\scas{vol}} = J^{1/3} \vec{I},
\eeq 

where $J=det(\vec{F})$ is the volume change, and $\vec{I}$ is the identity matrix. Consequently, the isochoric part of the right Cauchy Green deformation tensor is defined as 

\beq 
\label{eq2}
\vec{\bar{C}} = \vec{\bar{F}}^T\cdot \vec{\bar{F}}\, ,
\eeq 

with invariants 

\beq 
\label{eq3}
\bar{I}_1 = \vec{\bar{C}}:\vec{I}, \; \bar{I}_2 = \frac{1}{2}\left( (\vec{\bar{C}}:\vec{I})^2-\vec{\bar{C}}^2:\vec{I} \right).
\eeq 

Additionally, we introduce the vector $\vec{a}_0$ in the reference configuration which specifies the direction of anisotropy. This vector field gets mapped to the vector $\vec{a}=\vec{F}\cdot \vec{a}_0$ upon deformation. Since we deal with a two-dimensional domain, the fiber direction can be parameterized by the single angle $\theta$. The direction of anisotropy defines the fourth pseudo-invariant of $\vec{\bar{C}}$

\beq 
\label{eq4}
\bar{I}_4 = \vec{a}_0 \cdot \vec{\bar{C}} \cdot \vec{a}_0 \, . 
\eeq 

The collagen fibers are not necessarily perfectly aligned. In fact there is some dispersion around the preferred orientation $\vec{a}_0$. This dispersion can be understood as the result of fitting a uni-modal distribution to the fiber orientation distribution over a unit sphere \cite{gasser2005}. We use the parameter $\kappa$ to denote the dispersion, and we remark that this parameter is associated with the width of the orientation distribution. A value of $\kappa=0$ would mean all fibers being perfectly aligned, while $\kappa=1/3$ indicates uniform fiber distribution. Thus, rather than the value of the fourth pseudo-invariant alone, the fiber contribution consists of a linear combination of an isotropic term and an anisotropic term

\beq 
\label{eq5}
\bar{E} = \kappa (\bar{I}_1-3) + (1-3\kappa)(\bar{I}_4-1) .
\eeq

The strain energy can now be defined as a sum of a volumetric part and two isochoric parts, one that describes the isotropic ground matrix, plus the anisotropic term

\beq 
\label{eq6}
\Psi = \Psi^{\scas{vol}} + \bar{\Psi}^{\scas{iso}} + \bar{\Psi}^{\scas{f}}\, ,
\eeq 

with the volumetric term being

\beq 
\label{eq7}
\Psi^{\scas{vol}} = \frac{K}{2}\left(\frac{J^2-1}{2}-\ln{J}\right) \, ,
\eeq

controlled by the single parameter $K$ which is the bulk modulus. In this work, incompressibility is imposed fully and the parameter $K$ does not have any influence. The first contribution to the isochoric term is purely isotropic 

\beq 
\label{eq8}
\bar{\Psi}^{\scas{iso}} = \frac{\mu}{2}(\bar{I}_1 - 3) \, ,
\eeq 

and parameterized by the shear modulus $\mu$. The last term is due to the fiber family and takes the form 

\beq
\label{eq9}
\bar{\Psi}^{f} = \frac{k_1}{2k_2}\left(\exp{k_2 \langle \bar{E} \rangle^2-1} \right)
\eeq 

where $\langle \bullet \rangle$ denotes the Macaulay brackets, and the parameters $k_1$ and $k_2$ have been introduced. 

Thus, the material model is fully described by the inputs $[\mu,k_1,k_2,\kappa,\theta]$. Based on our previous work and a review of the literature, plausible parameter ranges are illustrated in Table \ref{table01} \cite{tonge2013,lee2019}. 

\begin{table}
    \centering
    \begin{tabular}{c|c}
         Parameter & Range  \\
         \hline 
         $\mu$ [MPa] & [0.004774, 0.006804] \\
         \hline 
         $k_1$ [MPa] & [0.0038, 0.2093] \\
         \hline 
         $k_2$ [-] & [52.53, 161.86] \\
          \hline 
          $\kappa$ [-] & [0.133, 0.333] \\
          \hline 
          $\theta$ [$^\circ$] & [0, 180]
    \end{tabular}
    \caption{Ranges for the parameters used to describe the mechanical behavior of skin including anisotropy}
    \label{table01}
\end{table}

\subsection*{Creating GP surrogates}
To create the surrogate model, the GOH parameter space summarized in Table 1 is sampled $N$ times using Latin hypercube sampling (LHS). LHS ensures that each parameter is sampled uniformly. Individual inputs are denoted as $x^{(n)}=(\mu^{(n)}, k_1^{(n)}, k_2^{(n)}, \kappa^{(n)}, \theta^{(n)})$ with a total training set
\beq
\label{eq10}
\mathbf{X}=(x^{(1)},...,x^{(N)})\in \mathbb{R}^{5 \times N}
\eeq
Each $x^{(n)}$ defines a FE simulation that creates a nodal strain output $E^{(n)}\in \mathbb{R}^M$ where $M$ represents the total number of free nodes in the mesh. This gives an overall strain matrix of
\beq
\label{eq11}
\mathbf{E} = (E^{(1)},...,E^{(n)})\in \mathbb{R}^{M \times N}
\eeq
For each row in the overall strain matrix $\mathbf{E}$, the outputs are centered by subtracting the row's mean and dividing by its standard deviation. This gives a centered and scaled output of
\beq
\label{eq12}
\mathbf{\bar{E}}=(\bar{E}^{(1)},...,\bar{E}^{(N)}) \in \mathbb{R}^{M\times N}
\eeq
Since the meshes have a large number of nodes, the strain outputs $E^{(n)}$ have a very high dimension $M$. Here, we reduce the dimensionality of this output data using principal component analysis (PCA) \cite{bishop2006pattern}. To perform a PCA, we define $W \in \mathbb{R}^{M \times M}$ as the linear transformation 

\beq
\label{eq13}
\vec{W}\vec{\bar{E}} = \vec{Y}
\eeq

where $\vec{Y} \in \mathbb{R}^{M \times N}$ is the principal component (PC) score matrix. In order to find this matrix $\vec{W}$, the singular value decomposition (SVD) of $\vec{\bar{E}}$ is introduced first,

\beq
\label{eq14}
\vec{\bar{E}} = \vec{U}\vec{\Sigma} \vec{V}^\top , 
\eeq
 
which consists of orthogonal matrices $\vec{U} \in \mathbb{R}^{M \times M}$ and $\vec{V} \in \mathbb{R}^{N \times N}$, and a rectangular diagonal matrix with non-negative real numbers $\vec{\Sigma} \in \mathbb{R}^{M \times N}$. Multiplying eq. (\ref{eq14}) from the left by $\vec{U}^{-1}$, and noting that because $\vec{U}$ is an orthogonal matrix $\vec{U}^{-1}=\vec{U}^\top$, we obtain

\beq
\label{eq15}
\vec{U}^{\top}\vec{\bar{E}}= \vec{\Sigma} \vec{V}^\top .
\eeq

Using eq. (\ref{eq15}) we define the PC projection in terms of the SVD: $\vec{W}=\vec{U}^\top$, $\vec{Y}=\vec{\Sigma} \vec{V}^\top$. The rows of $\vec{W}=\vec{U}^\top$ are the PCs, the rows of $\vec{Y}$ are the PC scores, or the projection of the original data (${\bf \bar{E}}$) to the PC basis. Since we centered the data prior to performing the SVD, the variance of the data is the square of the singular values scaled by $1/(N-1)$. Essentially, the singular values show how the information in each column of ${\bf \bar{E}}$ is distributed when we project it to the PC basis. Thus, we can truncate our PC values by defining a criterion, such as capturing 99$\%$ of the total variance in the data. This truncation leads to a reduced basis basis $\vec{W}' \in \mathbb{R}^{M'\times N}$ with PC scores $\vec{Z}=\vec{Y}'\in \mathbb{R}^{M'\times N}$. Ideally, the majority of the variance can be captured within a small number of PCs, meaning $M \gg M'$, significantly reducing the dimensionality of our initial nodal strain data. Indeed, as will be shown later, about thirty PCs capture most of the variance in our strain profiles. The GP regression is performed  on the truncated PC score data
\beq
\label{eq16}
\mathbf{Z} = (z^{(1)},...,z^{(N)})\in\mathbb{R}^{M'\times N} .
\eeq

The training dataset consists of the corresponding inputs paired with the truncated PC scores 
\beq
\label{eq17}
\mathcal{D}\equiv\{(x^{(n)}, z^{(n)})\}_{n=1}^N .
\eeq

Given the training data, we are interested in performing GP regression to learn a scalar function for each PC score. For any set of observations of the \textit{m}th PC score ($z_m\in\mathbb{R}^N$), we are interested in the function $f_m(x)$, but we consider that rather than observing the correct value of the function, we observe a noisy output

\beq
\label{eq18}
z_m^{(n)}=f_m(x^{(n)})+\varepsilon_m^{(n)}
\eeq

with $\varepsilon_m$ having independently and identically distributed Gaussian noise with zero mean and variance $\sigma^2_{ns,m}$. To learn the function $f_m(x)$ from the data, we first model the prior state of knowledge about $f_m$ using a GP
\beq
\label{eq19}
f_m(\cdot) \sim \mathcal{GP}(\mu_m(\cdot), k_m(\cdot))
\eeq
with mean and covariance functions $\mu_m(\cdot)$ and $k_m(\cdot)$, respectively. We choose a zero mean function such that the GP prior is a multivariate normal distribution over the inputs $\mathbf{X}$,
\beq
\label{eq20}
f_m(\mathbf{X}) \sim \mathcal{N}(0, \mathbf{K_m})
\eeq
The covariance matrix $\mathbf{K_m}\in\mathbb{R}^{N\times N}$ contains our assumption about the regularity of the function we wish to capture. Here we use a radial basis function as a kernel, and the components of the covariance matrix follow 
\beq
\label{eq21}
\begin{aligned}
K_{m(i, j)} &= k_m(x^{(i)}, x^{(j)}; \zeta_m)\hfill \\ &= s^2_{f, m} \exp{-\frac{1}{2}(x^{(i)}-x^{(j)})^\top \Lambda_m^{-1}(x^{(i)}-x^{(j)})}
\end{aligned}
\eeq
with $\Lambda_m = \mathrm{diag}(\lambda_{m, 1}, \lambda_{m, 2}, \lambda_{m, 3}, \lambda_{m, 4}, \lambda_{m, 5})$, and each $\lambda_{m, i}$ capturing the squared characteristic length-scale of each input. The process variance is denoted $s^2_{f, m}$ in eq (\ref{eq9}). These hyperparameters, denoted $\zeta_m$ in eq. (\ref{eq21}), are estimated by maximizing the likelihood of the observed outputs $z_m$,
\beq
\label{eq22}
\log{ p(z_m|X, \zeta_m)} \coloneq -\frac{1}{2}\vec{z}_m^\top \vec{\Sigma}_m^{-1}\vec{z}_m-\frac{1}{2}\log{|\vec{\Sigma}_m|}-\frac{N}{2}\log{2\pi} .
\eeq

Note that due to the assumption of the noisy observations we have a covariance  $\vec{\Sigma}_m=\vec{K}_m+\sigma^2_{ns, m}\vec{I}$, which is the sum of the GP covariance matrix plus the Gaussian noise. Having determined the hyperparameters $\theta_m$ from maximizing eq. (\ref{eq22}), the posterior of $f_m(x)$ is derived using Bayes' rule. Moreover, the posterior for any new parameter input $x^{(*)}$ is also Gaussian,
\beq
\label{eq23}
f_m(x^{(*)})|\mathcal{D}, x^{(*)}, \theta_m \sim \mathcal{N}(\mu_m(x^{(*)}; \zeta_m), \sigma^2_m(x^{(*)}; \zeta_m))
\eeq
 with predictive mean and variance,
\beq
\label{eq24}
\mu_m(x^{(*)};\zeta_m)=\vec{k}_m^\top \vec{\Sigma}_m^{-1}\vec{z}_m\, ,
\eeq
\beq
\label{eq25}
\sigma^2_m(x^{(*)};\zeta_m)=\vec{k}(x^{(*)}, x^{(*)}; \zeta_m)+\sigma^2_{ns,m}-\vec{k}_m^\top \vec{\Sigma}_m^{-1}\vec{k}_m \, .
\eeq
 
 In (\ref{eq23}) and (\ref{eq24}), the vector $\vec{k}_m$ is
 
\beq
\label{eq26}
\vec{k}_m=(k(x^{(*)},x^{(1)}; \zeta_m), ..., k(x^{(*)},x^{(N)}; \zeta_m)).
\eeq

In summary, eq. (\ref{eq23}) predicts the expected PC scores as the predictive mean (eq. \ref{eq24}), with error bars given by the predictive variance (eq. \ref{eq25}). 

To validate the GP model, we sample independently another set of $Q$ inputs using LHS

\beq
\label{eq27}
\vec{X}_v=(x_v^{(1)},...,x_v^{(Q)})\in \mathbb{R}^{5\times Q} .
\eeq

FE analyses are run for each of these Q input parameter sets for each flap to obtain truth data (the nodal strain outputs) to be compared against the GP predictions. Note that the GP predictions lie on the PC space. To obtain nodal strain outputs from the surrogates, we first use the posterior GP to obtain the predictive means of the PC scores

\beq
\label{eq28}
\vec{Z}_v=(z_v^{(1)}, ..., z_v^{(Q)}) \in \mathbb{R}^{M' \times Q}\, ,
\eeq

and then perform an inverse PCA transformation to obtain a centered prediction of the nodal strain as

\beq
\label{eq29}
\vec{\bar{E}}_v=\vec{W}'^\top \vec{Z}_v .
\eeq

Lastly, reversing the centering operation that was done before PCA in the training data, we obtain nodal strain predictions that can be compared directly to the FE truth.

\subsection*{Sensitivity Analysis}

To understand the influence of each parameter, we perform a Sobol sensitivity analysis \cite{sobol2001global}. The method followed in this manuscript follows closely our previous work and is also aligned with the documentation of Python's SALib library \cite{herman2017salib}. The goal of this analysis is to determine which of the five inputs has the largest influence on the strain distribution. Intuitively, if there is a large variation in the results of the surrogate model when one parameter is held constant while the others vary, that parameter likely does not have a large influence on the strain distribution. By contrast, if there is only a small variation in results when that parameter is held constant, it is likely influential. The sensitivity analysis thus requires a large number of function evaluations as different parameters are varied. The Satelli sampling scheme is used to obtain a total of $S\times(2K+2)$ samples, where $S$ is a large number, for example on the order of a thousand, and $K$ is the dimension of the input space. Clearly, this kind of analysis would be difficult to do with the original FE model. Instead, having trained the GPs, we can use the surrogates to perform the many function evaluations required for the sensitivity analysis. The Sobol sensitivity analysis on the outputs, obtained via GP evaluations, decomposes the variance in the model output into variance that can be attributed to each input and to the interaction between inputs. 

\subsection*{Optimization of surgical plan}
Another type of task that is enabled with the GP surrogates is flap optimization. As will be shown in the Results section, the fiber direction ($\theta$) is the most important parameter for the resulting strain distribution following reconstructive surgery.  While the surgeon does not control the fiber direction of the patient's skin, they do control how the flap is oriented with respect to the anisotropic features of skin, referred clinically as the relaxed skin tension lines \cite{borges1984relaxed}. Therefore, knowledge of the optimal fiber direction with respect to the orientation of the flap should be used for planning a surgery, even when other parameters are unknown or uncertain. We work to determine an optimal fiber direction through minimizing cost functions.

Here we propose three cost functions informed by clinical guidelines \cite{rohrer2005transposition,leedy2005reconstruction,2014GaSp}: (1) $\mathcal{C}_1$: mean nodal strain, (2) $\mathcal{C}_2$: sum of tensile strains at key points near the wound edges, (3) $\mathcal{C}_3$: sum of tensile strain at key points only at the distal ends of the flaps. With $x=(\mu, k_1, k_2, \kappa, \theta)$, mathematically, we have

\beq 
\begin{aligned}
\mathcal{C}_1(x) &= E_{\scas{avg}}, \; \\
\mathcal{C}_2(x) &= \sum_{\scas{Edge \, pts} \, i} \langle E_{i} \rangle, \; \\
\mathcal{C}_3(x) &= \sum_{\scas{Distal \, pts} \, i} \langle E_{i} \rangle,\\
\end{aligned}
\eeq 

where $\langle \bullet \rangle$ denote the Macaulay brackets. These cost functions depend on all the material parameters. However, as mentioned, only the anisotropy direction $\theta$ is a design parameter for surgery. Thus, we introduce the random vector $\phi = (\mu, k_1,k_2,\kappa)$ capturing all the material parameters other than $\theta$. 
Assuming risk neutrality, we should select $\theta$ by minimizing the expectation of the cost over the distribution of the parameters $\phi$ \cite{keeney1993decisions}:

\beq 
\label{eq_expected_cost}
\theta_{rn}^* = \argminA_{\theta} \mathbb{E}_{p(\phi)}[\mathcal{C}_i]\, .
\eeq 

To calculate the expectation we take advantage of the GP and simply draw many samples from $p(\phi)$ for a given value of $\theta$. In practice, $1000$ samples per $\theta$ are sufficient to obtain a sufficiently converged estimate of the expectation. The minimization with respect to $\theta$ is done by calculating the expectation of the cost for a sufficiently refined grid of $\theta$ values. In our case, we find that the function is smooth enough such that $181$ values of $\theta$ (all values from $0^\circ$ to $180^\circ$ by $1^\circ$, inclusive) are enough to interpolate the expected cost. After performing these calculations, we get the expectation for each value of $\theta$ and find the minimum. 

Under the assumption of extreme risk aversion, another strategy for determining an optimal fiber direction is to minimize the worst-case scenario. In this case the problem is:

\beq 
\label{eq_worst_cost}
\theta_{ra}^* = \argminA_{\theta} \argmaxA_{\phi \in \mathrm{supp}(p(\phi))} \mathcal{C}_i\,
\eeq

where $ \mathrm{supp}(p(\phi))$ is the support of the material parameter distribution. For the most general case, the support corresponds to the parameter range defined in Table \ref{table01}. The maximization problem with respect to $\phi$ is solved with a modified particle swarm optimizer \cite{shi1998modified}. The algorithm begins by generating $J$ random individual inputs of the form $\phi_{j}=(\mu_j, k_{1j}, k_{2j}, \kappa_j)$. These are the particles in the swarm. For each particle $j$, we then generate a random velocity ($v_j(t=0)$) and calculate its current objective value ($\mathcal{C}_j(t=0)$), with $t$ a pseudo-time used in the algorithm. These objective values are set as the initial personal best locations ($\hat{\phi}_j(t=0)$) for each particle. The best objective value of all the personal bests is the global best location ($g(t=0)$). After initialization, the iterations are attempts to adjust the location of the particles towards an optimal value. In each generation, or iteration, the velocity of each particle is updated based on
\beq
\label{eq31}
v_j(t+1)=\omega v_j(t) + c_1 r_1 [\hat{\phi}_j(t)-\phi_j(t)]+c_2 r_2 [g(t)-\phi_j(t)]
\eeq
Here, $\omega v_j(t)$ is the inertial component and updates the velocity using the influence of the particle's current direction of movement. $\omega$ is the inertial coefficient, generally taking a value between 0.8 and 1.2. The cognitive component, $c_1 r_1 [\hat{\phi}_j(t)-\phi_j(t)]$, updates the velocity based on the influence of the particle's previous best position. $c_1$ is referred to as the cognitive coefficient and generally takes a value of approximately 2, while $r_1$ is a random value between zero and one. The last portion of the velocity update is the social component, $c_2 r_2 [g(t)-\phi_j(t)]$. Again, $c_2$ is the social coefficient and generally takes a value of about 2 while $r_2$ is a random value between zero and one. This last component updates the velocity with respect to the current global best position. Overall, each particle's velocity is updated under the influence of its current direction of motion, previous best position, and the global best position. The particle's position is then updated as
\beq
\label{eq32}
\phi_j(t+1)=\phi_j(t)+v_j(t+1)
\eeq
If this new position is within the valid range for $\phi$ and has an improved objective value compared to the current personal best, it becomes the particle's personal best. The global best for the next generation is updated from the new positions at the end of the iteration. The algorithm continues until the maximum number of generations is reached or the global best solution fails to improve for 5 consecutive generations. The overall result of the optimization is then the global best position and its corresponding objective function value. This optimization is repeated 5 times to account for the stochasticity in the algorithm and ensure that we do not move forward with a sub-par optimization result. 

Similar to the minimization problem in (\ref{eq_expected_cost}), for (\ref{eq_worst_cost}) we interpolate the worst cost as a function of $\theta$ by solving the maximization over $\phi$ for $181$ values of $\theta$. From this, we are then able to determine the value for theta that minimizes this worst case scenario.

Finally, a brief discussion on $p(\phi)$ is needed. For the most general case, we perform the optimization of the flap using the prior of the material parameters $p_0(\phi)$, which is a uniform distribution over the entire range in Table \ref{table01}. This might be too broad of a range. Hence, we also consider a distribution for the material parameters based on one of the patient's values reported in \cite{tonge2013}.  

\begin{figure*}[!htb]
\begin{center}
\includegraphics[width=0.8\textwidth]{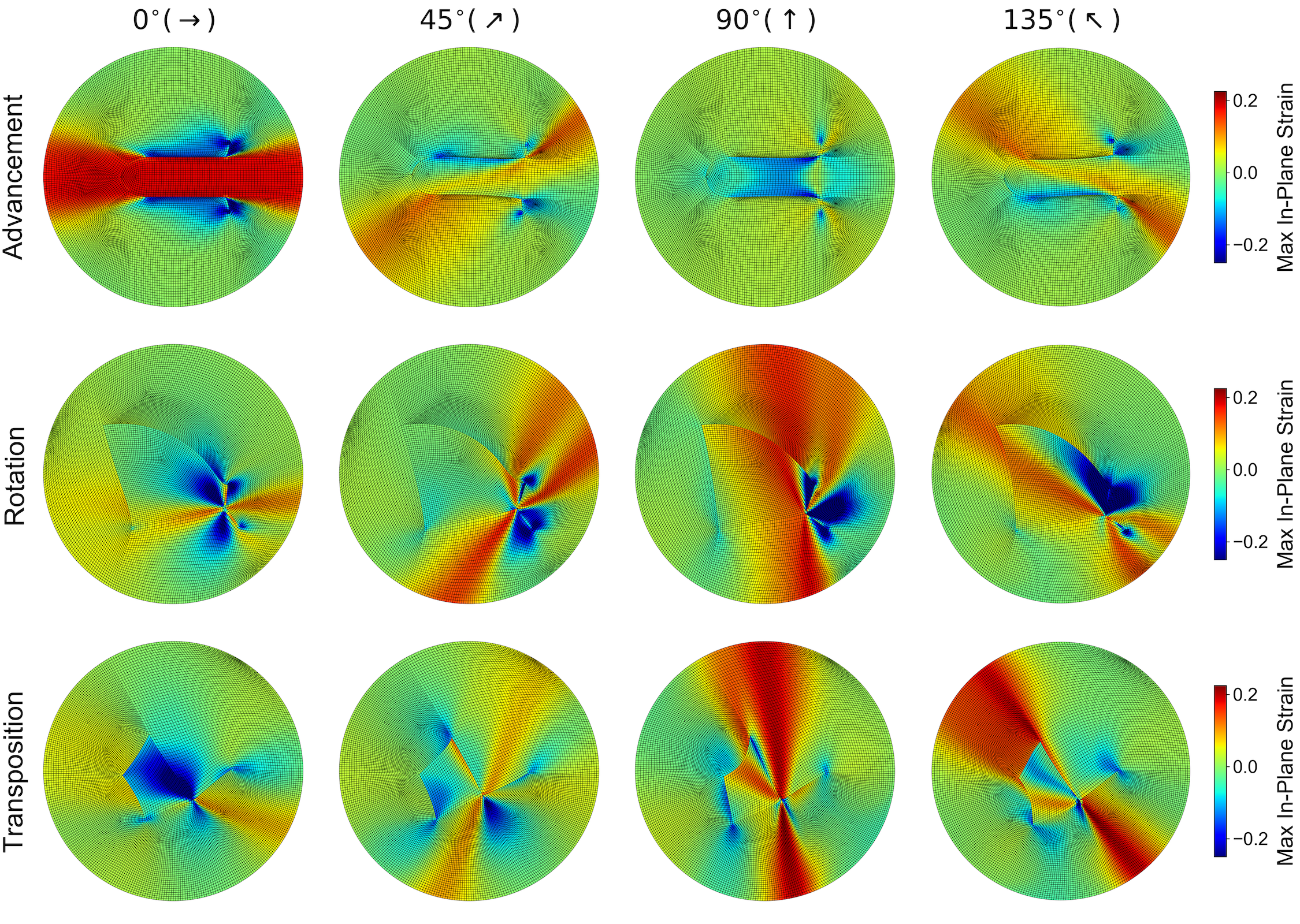}
\caption{The distribution of maximum in-plane strain for the advancement, transposition, and rotation flaps at $\theta = [0^\circ, 45^\circ, 90^\circ, 135^\circ]$ with all other parameters at mean values of their range. Strain patterns show features spanning the entire skin region and aligning with the fiber orientation.}
\label{fig02}
\end{center}
\end{figure*}

\section*{Results}
\subsection*{Exploring the effect of anisotropy in advancement, rotation and transposition flaps}
To begin, we explored the impact of anisotropy on strain profiles produced from FE simulations. We ran 12 simulations taking the mean values for $\mu$, $k_1$, $k_2$, and $\kappa$ from the range in Table \ref{table01} and varying $\theta$ at $[0^\circ, 45^\circ, 90^\circ, 135^\circ]$. The results are shown in Figure \ref{fig02}, where it can be observed that the overall trend of the strain profiles follows the fiber direction. While these results are only true for one set of parameters $[\mu,k_1,k_2,\kappa]$, they support the intuition that direction of anisotropy plays an important role in flap design. For instance, for the advancement flap, fibers aligned with the direction of advancement lead to high strains across the flap and surrounding skin, from the proximal to the distal ends. As the fiber is rotated to $45^\circ$ with respect to the flap, the higher strains are oblique with respect to the advancement direction. Interestingly, fibers oriented perpendicular to the direction of advancement lead to compressive strains at the middle of the flap and overall small strains throughout the skin patch. The last simulation shown for advancement, $\theta=135^\circ$, is similar to $\theta=45^\circ$ since the flap is symmetric with respect to the horizontal axis.

In the rotation flap, fibers at $\theta = 0^\circ$ result in relatively small strains near the base of the flap, with a  couple of small regions under compression and a narrow bad of tissue in tension. In this case, the distal end has even lower strains. Rotating the fibers to $\theta = 45^\circ$ results in strains being greater along that direction, and overall increasing in magnitude. This trend continues as the fibers are rotated to $\theta = 90^\circ$. For $\theta = 135^\circ$, a band of high tensile strains extends from the base of the flap to the distal end. As opposed to the advancement flap, the rotation design does not have any symmetry. Additionally, the strain profiles from the rotation flap are more intricate with respect to advancement, particularly near the base of the flap, due to the point at which three different edges come together in a $Y$ junction. 

In the transposition case there are also no planes of symmetry, and changing the fiber gradually from $\theta = 0^\circ$ to $\theta = 135^\circ$ yields vastly different strain contours. The trends in the transposition flap are overall similar to the other designs, with higher strains aligned with the fiber direction. Interestingly, the transposition design entails greater reorientation of tissue, which can be seen as a more discontinuous pattern of strain across the suture lines in the deformed configuration. For example, when $\theta = 135^\circ$, the bottom-right region of tissue is rotated approximately $60^\circ$ before being sutured to the top-left edge. Due to the mismatch in fiber direction in the deformed state between the surrounding tissue and the distal portion of the flap that has been rotated, the surrounding skin at the top-left region is under high tensile strain while the flap immediately adjacent is actually not subjected to higher strains.

The magnitude of the strains are different across flaps for a given orientation. For instance, when $\theta = 90^\circ$, the advancement flap shows the smallest strains while the transposition flap has the highest strains. On the other hand, for $\theta = 0^\circ$, the advancement flap results in the highest strains compared to the other flaps. We remark again that the results in Figure \ref{fig02} were obtained by fixing all other parameters to their mean. In order to conduct a rigorous sensitivity analysis over the entire parameter range, the surrogate models are needed to efficiently sample the input space.

\begin{figure*}[!htb]
\begin{center}
\includegraphics[width=0.8\textwidth]{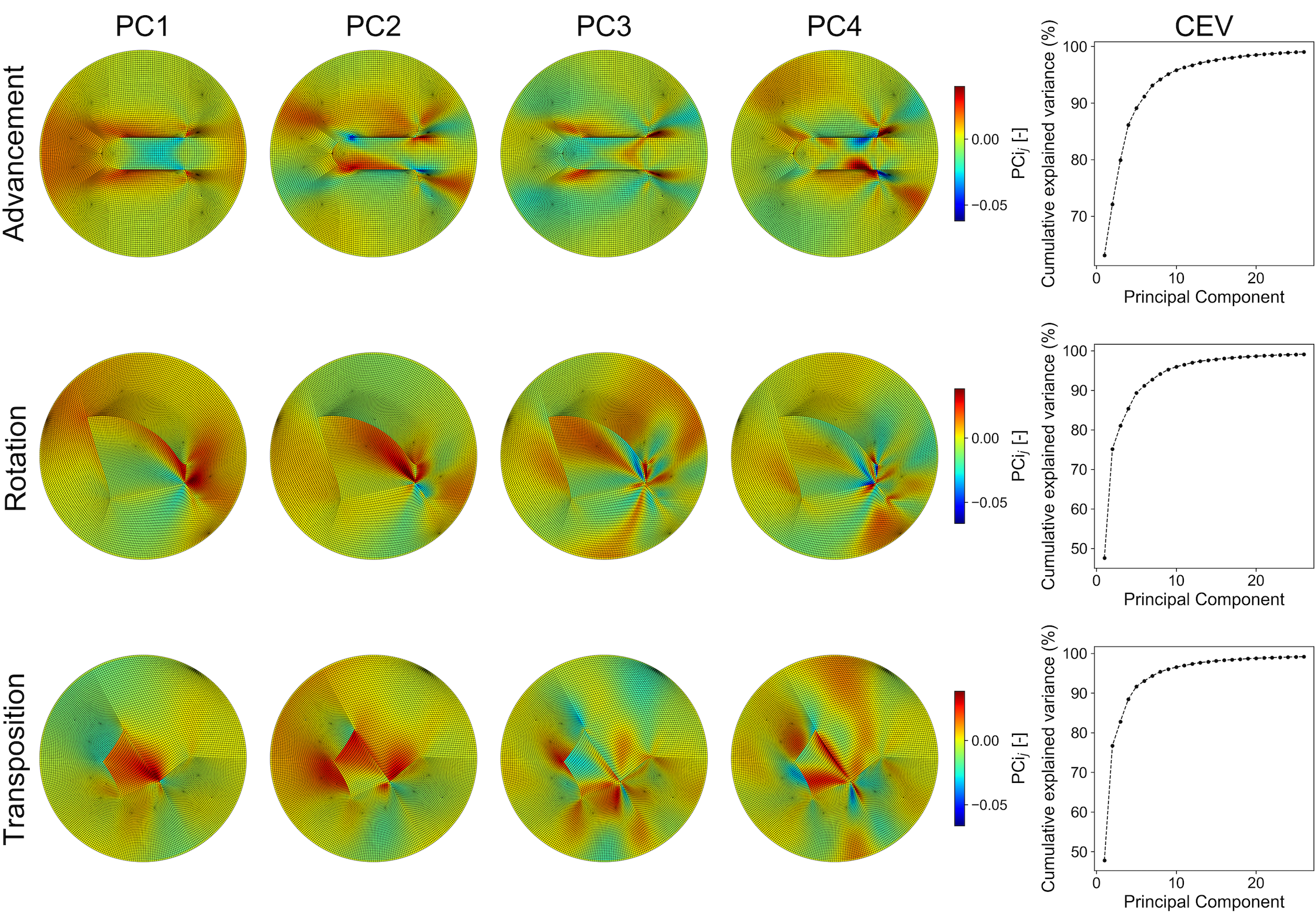}
\caption{First 4 PCs plotted on the finite element meshes for each of the three flaps. Cumulative explained variance (CEV) in accounting for 99\% of the variation in the total data is shown in the last column for each of the flaps. The first 4 PCs explain 86.1\%, 85.3\%,and 88.5\% of the advancement, rotation and transposition flaps respectively. PCs form an alternative basis for the strain profiles which enable compressing of the data into very few features compared to the number of nodes. Ultimately, to account for 99$\%$ of the variance, 26 PCs were kept as a truncated basis for the advancement case, 24 for the rotation, and 23 for the transposition flap.}
\label{fig03}
\end{center}
\end{figure*}

\subsection*{Creating and validating GP surrogates}
The surrogates were trained using $N=2000$ FE simulations and validated on an additional 400 parameter sets. The training and validation sets were obtained from separate LHS instances. The majority of the simulations ran without artificial energy dissipation. However, for the advancement case, a total of 159 simulations were run with some dissipation, 55 of those incurred in up to 0.0001\% artificial energy with respect to total energy, 63 dissipated up to 0.001\% , 30 simulations required 0.01\%, 10 more needed 0.1\% and only one was done with up to 1\% dissipated energy with respect to total energy. For the rotation flap, 205 simulations allowed for 0.001\% energy dissipation, 118 allowed 0.01\%, 8 had 0.1\%, and 6 were run with up to 1\% of artificial energy compared to total energy. For the transposition flap, out of the total, in 230 cases the simulation was allowed to dissipate up to 0.001\%, in 41 cases the maximum was capped at 0.01\%, and in 10 occasions there was an allowed maximum energy dissipation of 0.1\% .

After obtaining FE results for the strain distributions for each parameter set, we reduced the dimensionality of these data with PCA. Initially, the overall strain matrices had dimensions $2000 \times 12004$, $2000 \times 12218$, and $2000 \times 12389$ for the advancement, rotation, and transposition flaps, respectively. Using the PCA while retaining 99\% of the variation in the data, these were respectively reduced to $2000\times 26$, $2000 \times 24$, and $2000 \times 23$ datasets. The first 4 PCs for each flap alongside the cumulative variance explained with each additional PC are reported in Figure \ref{fig03}. Unfortunately, PCs do not necessarily entail any physical meaning or intuition, as opposed to the strain fields in Figure \ref{fig02}. It is worthwhile to point out that in previous work, ignoring anisotropy, less than 10 PCs were enough to account for $99.9\%$ of the variance in stress profiles of the same three flap designs \cite{lee2019}. Thus, even though anisotropy direction and fiber dispersion are only two additional parameters with respect to previous work, they contribute to more complex strain and stress distribution over the entire skin patches, clearly seen in Figure \ref{fig02}. Nonetheless, PCA is still able to reduce the dimensionality effectively.

\begin{figure*}[!htb]
\begin{center}
\includegraphics[width=0.8\textwidth]{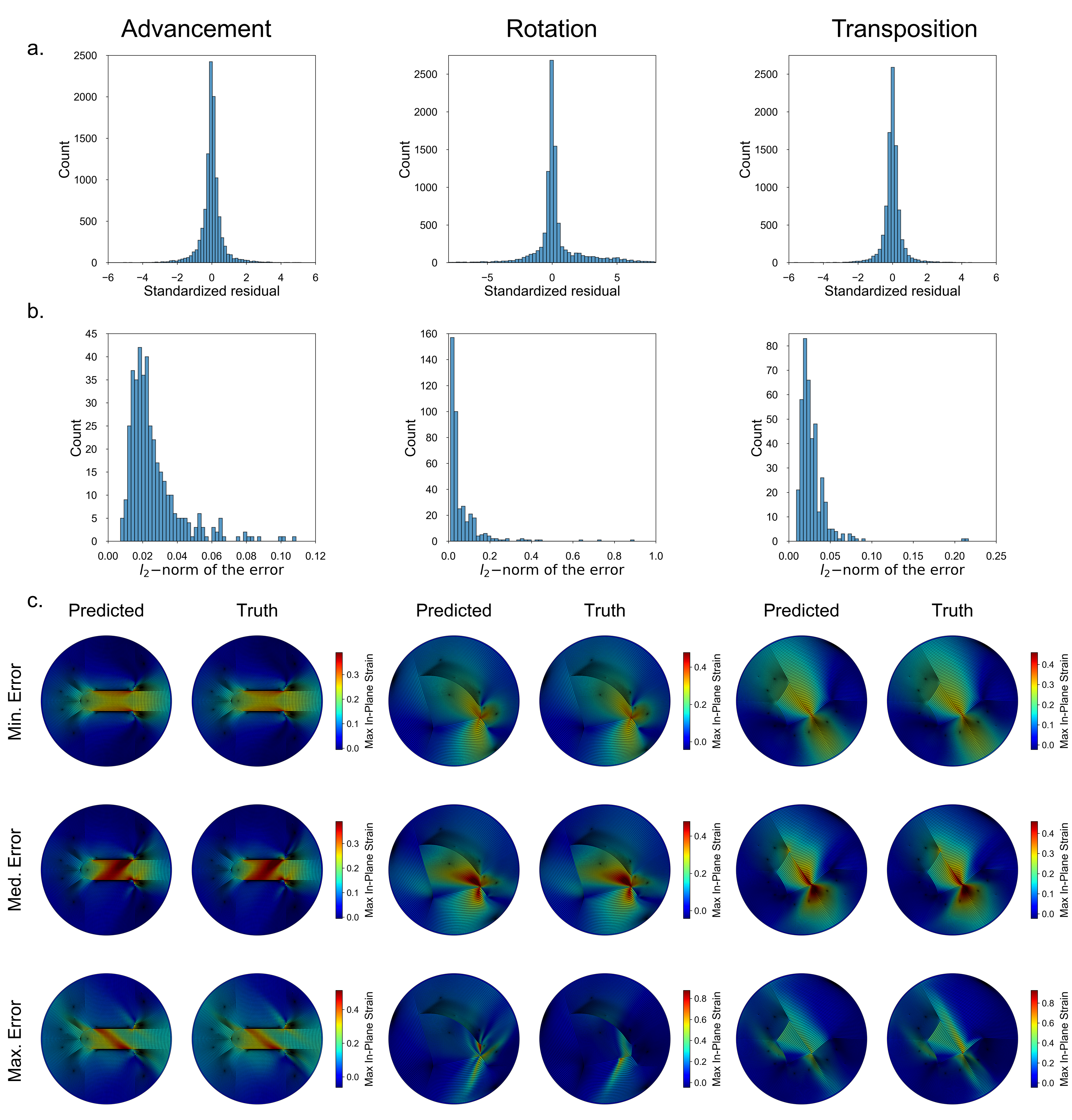}
\caption{a) Standardized residuals computed from comparing surrogate predictions and validation data after it has been projected onto the PC basis. b) L2 Norm of the relative error between the surrogate and the validation data. To obtain strains, the predictions of the PC scores from the individual GP surrogates are transformed via inverse PCA to the strain space. c) surrogate predictions versus FE truth for the minimum, median, and maximum L2 norm relative errors.}
\label{fig04}
\end{center}
\end{figure*}

We used the reduced data after truncating the PC basis at $99\%$ of the variance, and we trained independent GPs for each PC score for each flap. Next, we evaluated the quality of the surrogate models.  The validation simulations provide strain profiles and not PC scores. While indeed we are ultimately interested in the prediction of strains, first we investigated the performance of the individual GPs on the PC space. The strain profiles from the validation set were hence projected to the PC basis. Once the validation data was projected onto the reduced basis, we computed the standardized residuals for the predicted PC scores. The standardized residuals measure the difference between the prediction of the GP and the true value from the validation set, divided by the predicted variance of the GP. Results are plotted in Figure \ref{fig04}a. These histograms aggregate all the PC scores, i.e. we do not separate into individual PC scores for each flap. For all flaps and all PC scores, the standardized residuals fall mostly within the [-3, 3] confidence interval and are centered around zero. In other words, we know that the majority of the validation points fall within the 99.7\% confidence range predicted by the GP surrogates. However, some points do lie outside this range, especially for the rotation flap. This indicates that there may be areas of the parameter space that require further exploration, particularly for some of the PC scores of the rotation flap. 

The PC score prediction is not necessarily an indication of the performance of the surrogate on the strain space, which is the quantity of interest. To compare directly the prediction of the surrogates to the data from the validation set, we projected the PC score prediction to the strain space using the inverse PCA transformation. Next, we examined the L2 norm relative error of the strain prediction with respect to the validation set. Figure \ref{fig04}b depicts histograms of the L2 norm relative error and Figure \ref{fig04}c showcases comparisons between truth and prediction for the best, median, and worst L2 norm relative error cases for each flap. While the transposition and rotation flaps do include a few outliers in terms of the L2 norm, the vast majority of relative error values fall below 0.1 for advancement and transposition flaps and 0.2 for the rotation flap. For the advancement flap, the maximum error is 0.109 with a mean error of 0.026 and 99.5\% (398 of 400 validation simulations) of the errors falling below 0.1.  The transposition flap has a maximum error or 0.216 and mean error of 0.028 with 99.25\% (397 of 400 validation simulations) of the errors falling below 0.1. The rotation flap, however, has a wider distribution of error with a maximum value of 0.895, but with a mean of 0.065. In the rotation case, 82.0\% of the errors fall below 0.1 (328 of 400 validation simulations). We remark that even with the wider error distribution, even the worst prediction shows qualitative agreement with the truth as seen in Figure \ref{fig04}c, middle columns. 

\begin{figure*}[!htb]
\begin{center}
\includegraphics[width=0.8\textwidth]{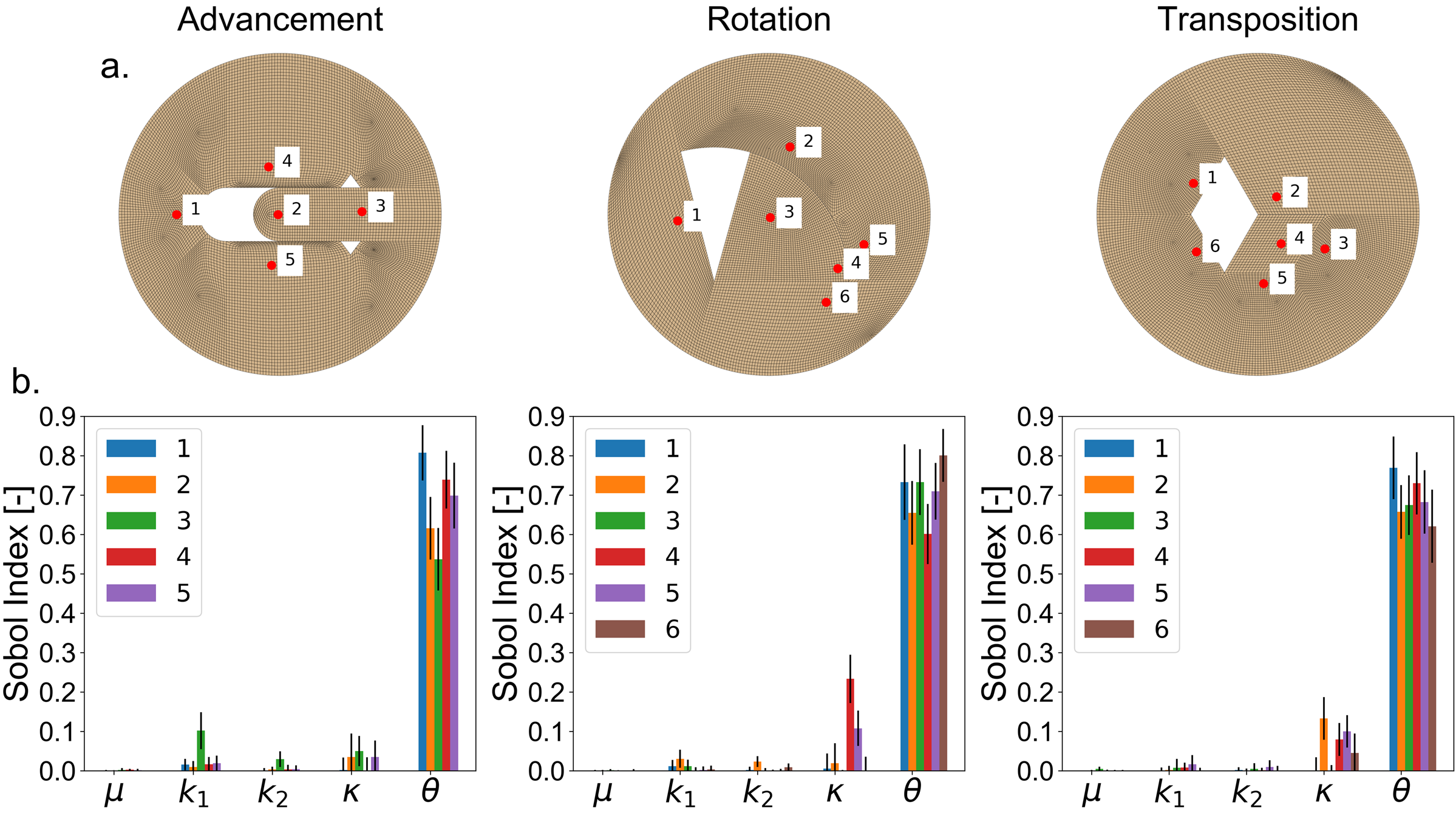}
\caption{a) Locations of key points used in the sensitivity analysis and in the optimization steps. b) Sobol index for the nodal strain at each of the key points after completing the analysis using 12,000 predictions from the surrogate model.}
\label{fig05}
\end{center}
\end{figure*}

\subsection*{Sobol Sensitivity Analysis}
Following the Methods section, we chose $S=1000$ for the sensitivity analysis, and sampled a total of $1000*(2(5)+2)=12000$ parameter sets using the Satelli sampling scheme. For each of these inputs, we used the surrogate models to predict nodal strains. Rather than looking at the entire strain field and defining a global scalar quantity of interest, we decided to focus on the strain value at specific locations. We chose points deemed more important indicators for the clinical setting \cite{rohrer2005transposition,leedy2005reconstruction}. These points are chosen near the suturing region, which is the zone where wound healing must take place, and that may be affected by the stress or strain more directly \cite{gurtner2011improving,buchanan2016evidence}. The points of interest are indicated in Figure \ref{fig05}a. The results from the Sobol sensitivity analysis using as quantity of interest the strain at the specific points are depicted in Figure \ref{fig05}b. These plots clearly illustrate that the fiber direction ($\theta$) has the most significant impact on variation in the nodal strains. As opposed to the results in Figure \ref{fig02}, which show strain variations for four different angles but fixing all other parameters, here we are able to sample the entire input space efficiently using the surrogate model.

\subsection*{Optimizing Flap Orientation}
In the clinical setting, the surgeon may not have access to reliable data on the material properties, particularly the inputs $\mu,k_1,k_2,\kappa$ to the model, which can only be obtained through mechanical testing \cite{jor2011,tonge2013}. However, surgeons do typically have knowledge of the anisotropy direction from anatomy and visual and manual inspection \cite{langer1978anatomy, borges1984relaxed}. Therefore, we decided to solve the optimization problem using the surrogate in order to find the best flap orientation with respect to the direction of anisotropy. As stated previously, excessive tension and deformation near a wound or sutured region is a cause for wound healing complications and pathological scarring \cite{gurtner2011improving,wong2011mechanical}. As such, the cost functions defined here generally focus on quantifying strain near the sutured regions. As defined previously, these functions are: i) mean nodal strain, ii) sum of tensile strains at key points near the would edges, and iii) sum of tensile strains at key point(s) only at the distal ends of the flaps. For the second cost function, all points selected for the sensitivity analysis are used. For the third cost function, point 2 is selected for the advancement flap, points 3 and 4 are selected for the rotation flap, and point 4 is selected for the transposition flap. Assuming that $\theta$ is the only input that can be controlled and that no other information is available regarding $\phi=[\mu,k_1,k_2,\kappa]$, we first considered that the prior assumption for the mechanical response $p_{0}(\phi)$ which is just a uniform distribution in the range of Table \ref{table01}. Additionally, we consider two smaller, normal distributions of parameters: one surrounding the 59 year old female parameters defined in \cite{tonge2013} and one surrounding the mean of the parameter ranges in Table \ref{table01}. The normal distributions here are defined to encompass  $\pm 10$\% of the given parameter values within 3 standard deviations on each side of the mean.

\begin{figure*}[!htb]
\begin{center}
\includegraphics[width=0.9\textwidth]{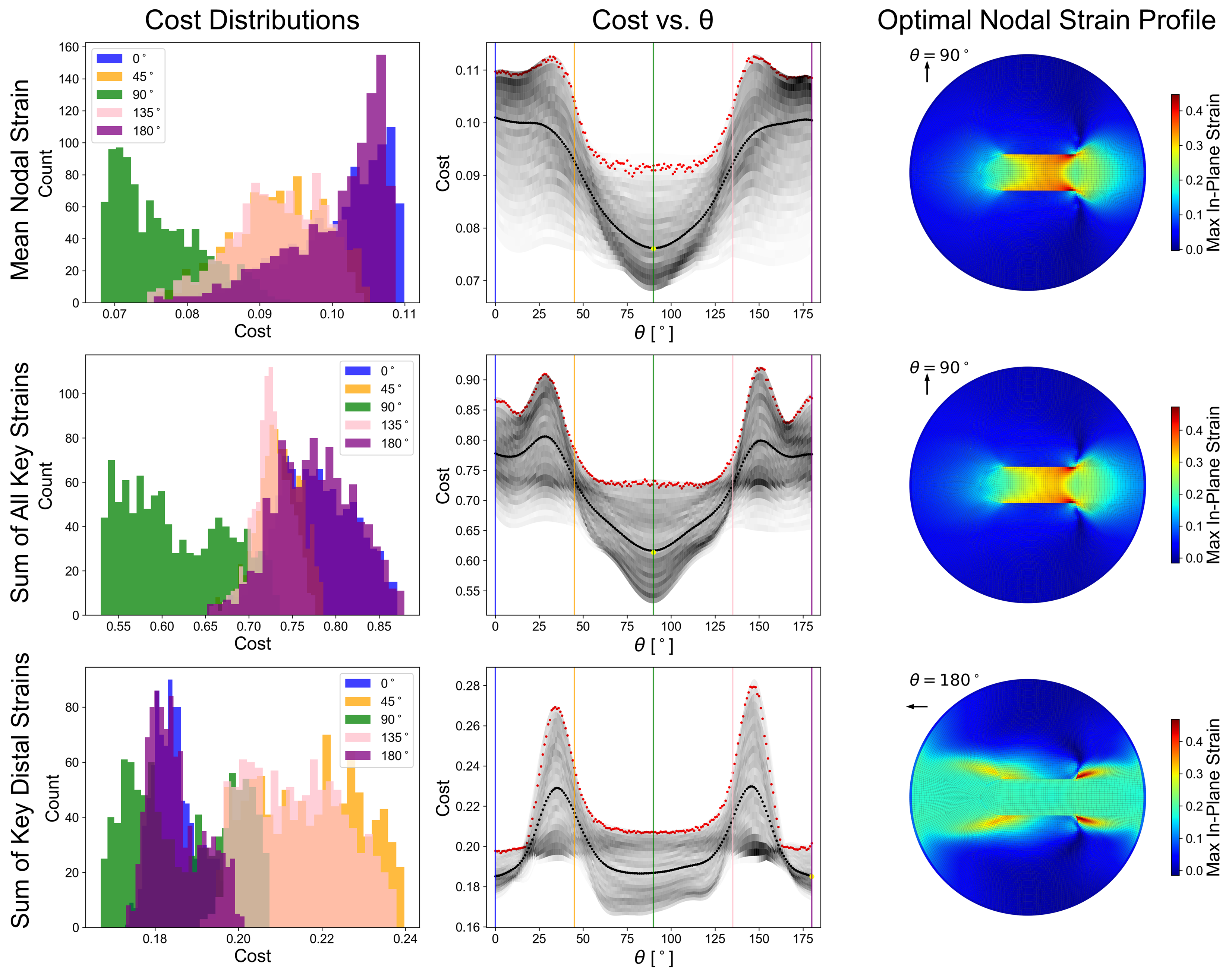}
\caption{Results of optimization for the advancement flap. The first column shows cost distributions for five values of $\theta$. The second column shows the cost versus $\theta$ plots. The vertical lines of different color correspond to the selected values of $\theta$ in the first column. In the middle column, the shaded area corresponds to the distribution of the cost as samples from $p_0(\phi)$ are taken. The expected value of the cost is shown as the black solid line, and the minimum of this expectation is the yellow point in the plot. The red data points show the worst case scenario computed from the particle swarm optimization for each $\theta$. The strain profile corresponding to the minimum expected cost is shown in the third column. The rows denote the three different cost functions introduced in the main text}
\label{fig06}
\end{center}
\end{figure*}

\begin{figure*}[!htb]
\begin{center}
\includegraphics[width=0.9\textwidth]{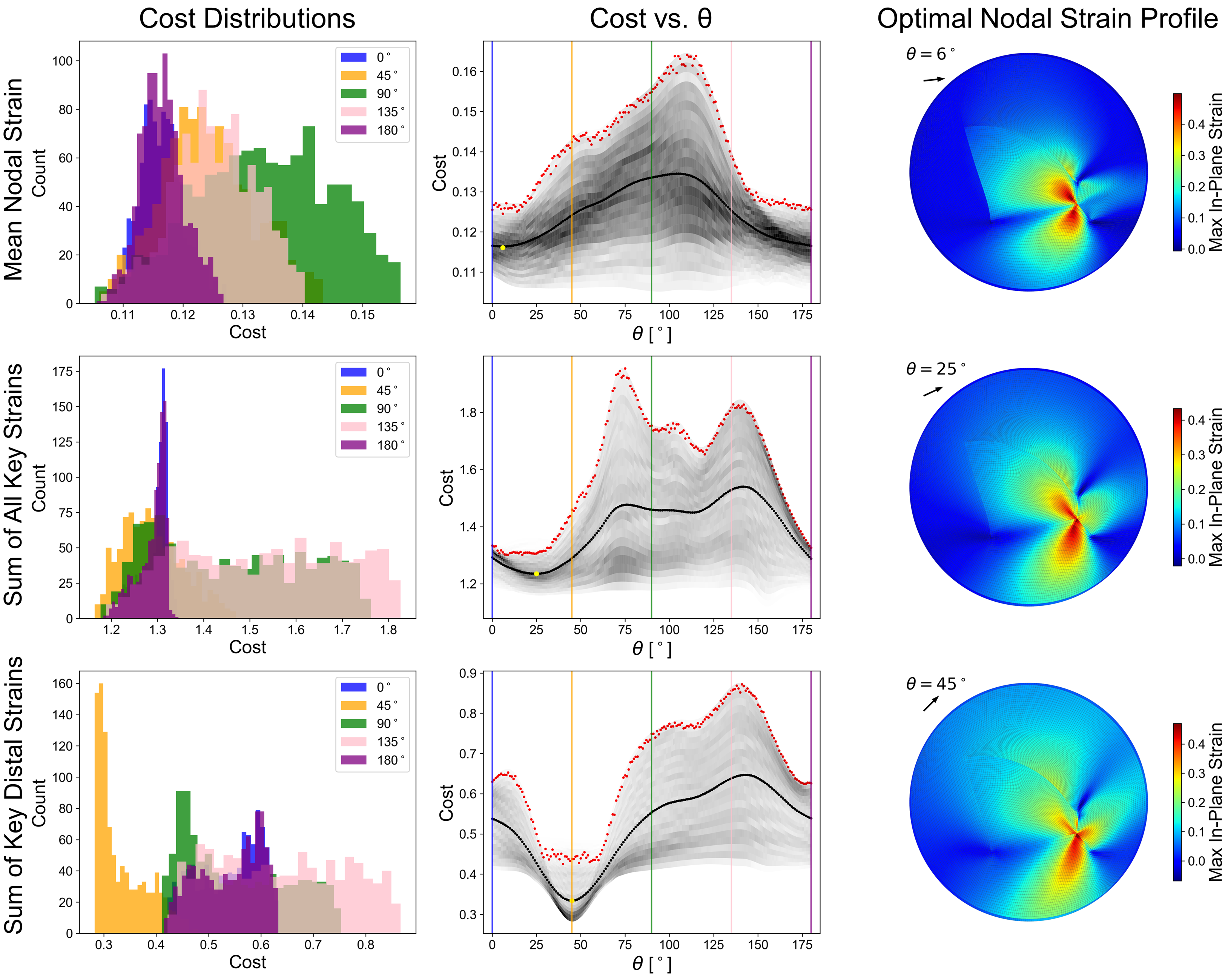}
\caption{Results of optimization for the rotation flap. The first column shows cost distributions for five values of $\theta$. The second column shows the cost versus $\theta$. The vertical lines  correspond to the selected values of $\theta$ in the first column. The shaded area shows the distribution of the cost obtained from sampling values from $p_0(\phi)$, the prior over the material parameters. The solid black line is the expected cost with minimum indicated by the yellow point. The red data points show the worst case scenario obtained from the particle swarm optimizer for each value of $\theta$. The strain profile corresponding to the minimum expected cost is shown in the third column. Each row corresponds to a different cost function.}
\label{fig07}
\end{center}
\end{figure*}

\begin{figure*}[!htb]
\begin{center}
\includegraphics[width=0.9\textwidth]{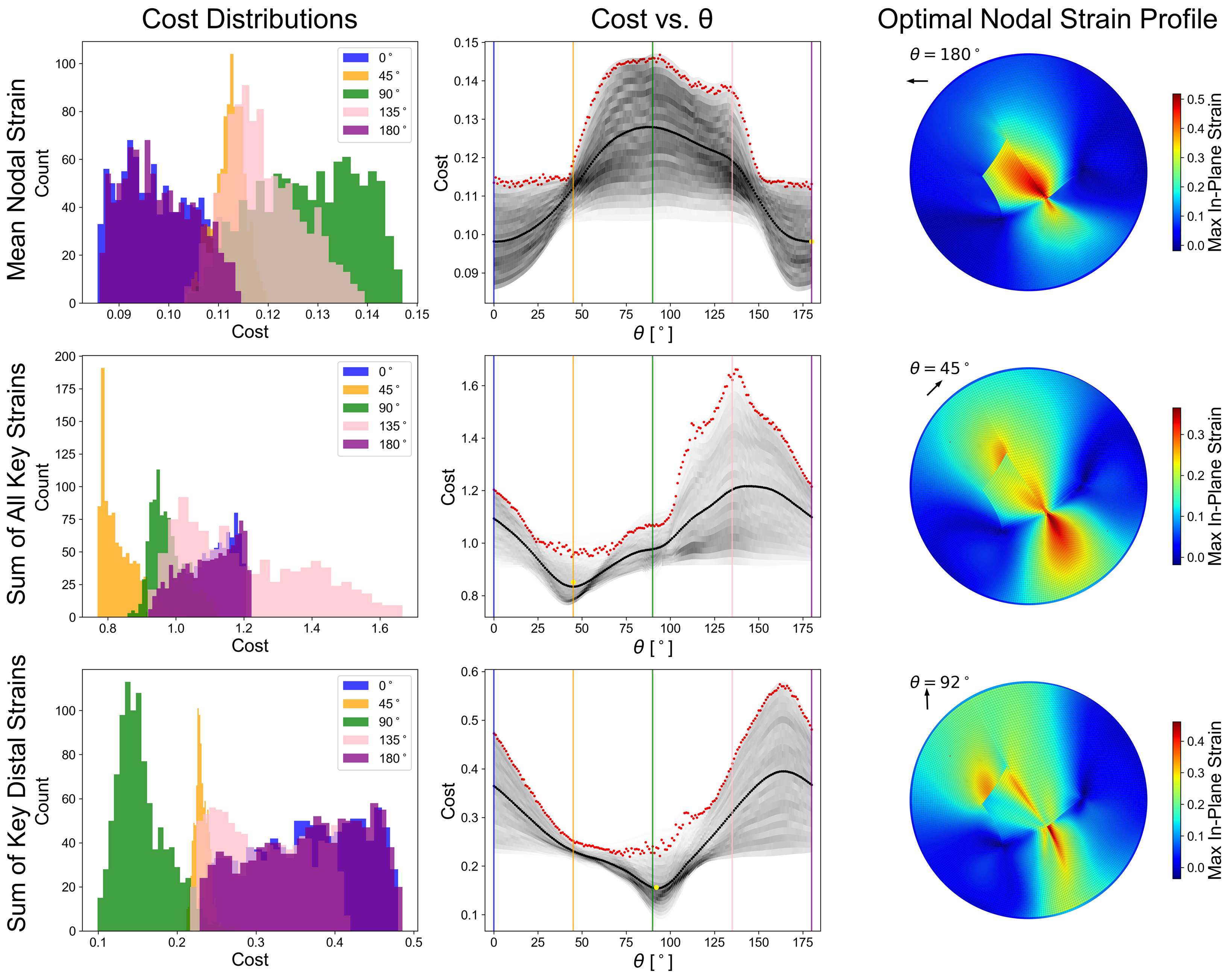}
\caption{Results of optimization for the transposition flap. The first column shows cost distributions for five values of $\theta$. The second column shows the cost versus $\theta$. The vertical lines of different color correspond to the selected values of $\theta$ in the first column. The solid black line in the middle column is the curve of the expected cost as a function of $\theta$, while the gray shaded regions show the distribution of the cost which follows from sampling material parameters from $p_0(\phi)$.The yellow points on the expected cost curve denotes the minimum of the curve. The worst case scenario for each $\theta$ obtained with the particle swarm optimizer is depicted with red points on the middle column. The strain profile corresponding to the minimum expected cost is shown in the third column.}
\label{fig08}
\end{center}
\end{figure*}

Using the methods described above, we complete both optimization problems - minimizing the expectation of the cost and minimizing the worst case scenario. The particle swarm optimization for $\phi$ was completed with 50 particles, a maximum of 25 generations, and optimizer hyperparameters $\omega = 1.2$, $c_1 = 2.0$, and $c_2 = 2.0$. The initial particle distribution was created by selecting 50 random values in the ranges listed in Table \ref{table01} and setting their initial velocities to random values between $0$ and $80\%$ of the range of values for each parameter. The results for the optimization considering the distribution $\phi_0$ are illustrated in Figures \ref{fig06}, \ref{fig07}, and \ref{fig08} for the advancement, rotation, and transposition flaps, respectively. These plots first show the cost distributions for $\theta = [0^\circ, 45^\circ, 90^\circ, 135^\circ]$. The second column shows the cost versus $\theta$ for all 181 values of $\theta$. The vertical lines correspond in color to the $\theta$ values selected to illustrate the cost distribution. The gray shaded region illustrates the distribution of the cost for each $\theta$. Darker shading indicates higher frequency as values of the cost are sampled by sampling the distribution $p_0(\phi)$. The black data points in the middle columns of Figures \ref{fig06}-\ref{fig08} are the expecteation of the cost for every $\theta$. The red data points in the Figures show the worst case scenario from the particle swarm optimization for each $\theta$. It is evident that, even though there is some variation, in general both the expectation of the cost and the worst case scenario follow the same trend with respect to $\theta$. Consequently, the solutions to both the optimization problems are approximately equivalent. The yellow points in the middle plots signal the value of $\theta$ for which the expected cost achieves the minimum. The last column shows the strain profile as predicted by the surrogate for the optimal points with respect to each of the cost functions.

For the advancement flap, the optimal orientation of the flap with respect to the anisotropy direction can be one of two options depending on which cost function is used. Minimizing the mean nodal strain or the sum of the strains on the key points around the suture line leads to the same optimal $\theta=90^\circ$. This result makes sense and follows clinical guidelines \cite{maciel2013local,buchanan2016evidence}, which recommend advancing perpendicular to the fiber orientation. When only the strain at a single point on the distal end is considered, the cost function is relatively flat and low around $\theta=90^\circ$, but there is also a local minimum around $\theta=0^\circ$. We plot this value in Fig. \ref{fig06} to illustrate the difference, although in reality, $\theta=90^\circ$ also achieves a low value of the expectation of the cost function. On the other hand, the distribution of the cost for a given $\theta$ suggests that the worst case scenario for may be worse for $\theta=90^\circ$. The strain profile for this last cost function (minimizing only the strain at one point in the distal end of the flap) shows higher strains overall, compared to optimization based on the other two cost functions. For comparison, side by side strain profiles for $\theta=90^\circ$ and $\theta = 180^\circ$ can be found in the supplement. We also remark that the cost function is symmetric around $\theta=90^\circ$ or $\theta=0^\circ$ which is expected since the flap design is symmetric.


Next, looking at the rotation flap (Fig. \ref{fig07}), we see three different optimal values for $\theta$: $6^\circ$, $25^\circ$, and $45^\circ$ for mean nodal strain, sum of all key strains, and sum of key distal strains, respectively. We no longer have the symmetry seen with the advancement flap, as the rotation flap is not symmetric across either major axis. We do, however, note that the optimal results for both objective functions are roughly equivalent as seen in the last column of Fig. \ref{fig07}. For all three functions, there is a wide distribution of the cost because we used the prior $p_0(\phi)$. Even though we will introduce narrower distributions, using the prior is still advantageous because it can provide general guidelines even when no specific patient information is available. One of the key insights from using this wide distribution is the dependence of the worst case scenario as a function of $\theta$. Looking at the red points in the middle column of Fig. \ref{fig07} it is clear that worst case scenario can vary widely as $\theta$ changes. Similar to the advancement case, the worst case scenario follows closely the trends from the expected cost, but offers perhaps a better argument to restrict $\theta$ to a smaller range of approximately $[10,40]$ even when no other information is available. 

For the transposition flap (fig. \ref{fig08}), we again do not have symmetry with respect to $\theta$. The different cost functions with respect to $\theta$ show unique features and lead to very different optimal strain profiles. The optimal values for $\theta$ are $180^\circ$, $45^\circ$, and $92^\circ$ for the mean nodal strain, sum of all key strains, and sum of key distal strains cost functions, respectively. Note that the cost functions are periodic and thus, for the mean nodal strain, the strain for $\theta=0^\circ$ is roughly equivalent to that for $\theta=180^\circ$. The worst case scenario in this flap also follows the trends from the expectation of the cost. The fact that the cost functions produce different results underscores the need to narrow down the design criteria for this flap. As mentioned in the Methods section, clinical guidelines support the idea of minimizing the overall deformation, but also suggest that points near the wound might be more at risk of complication such as wound dehiscence or scarring, while distal points in the flap may be more at risk of ischemia and necrosis \cite{barnhill1984study,gurtner2011improving,buchanan2016evidence}. The results shown here suggest that each of these objectives can lead to a different choice of $\theta$ and therefore a careful examination of the relative importance of each of these objectives is needed.  

\begin{figure*}[!htb]
\begin{center}
\includegraphics[width=0.9\textwidth]{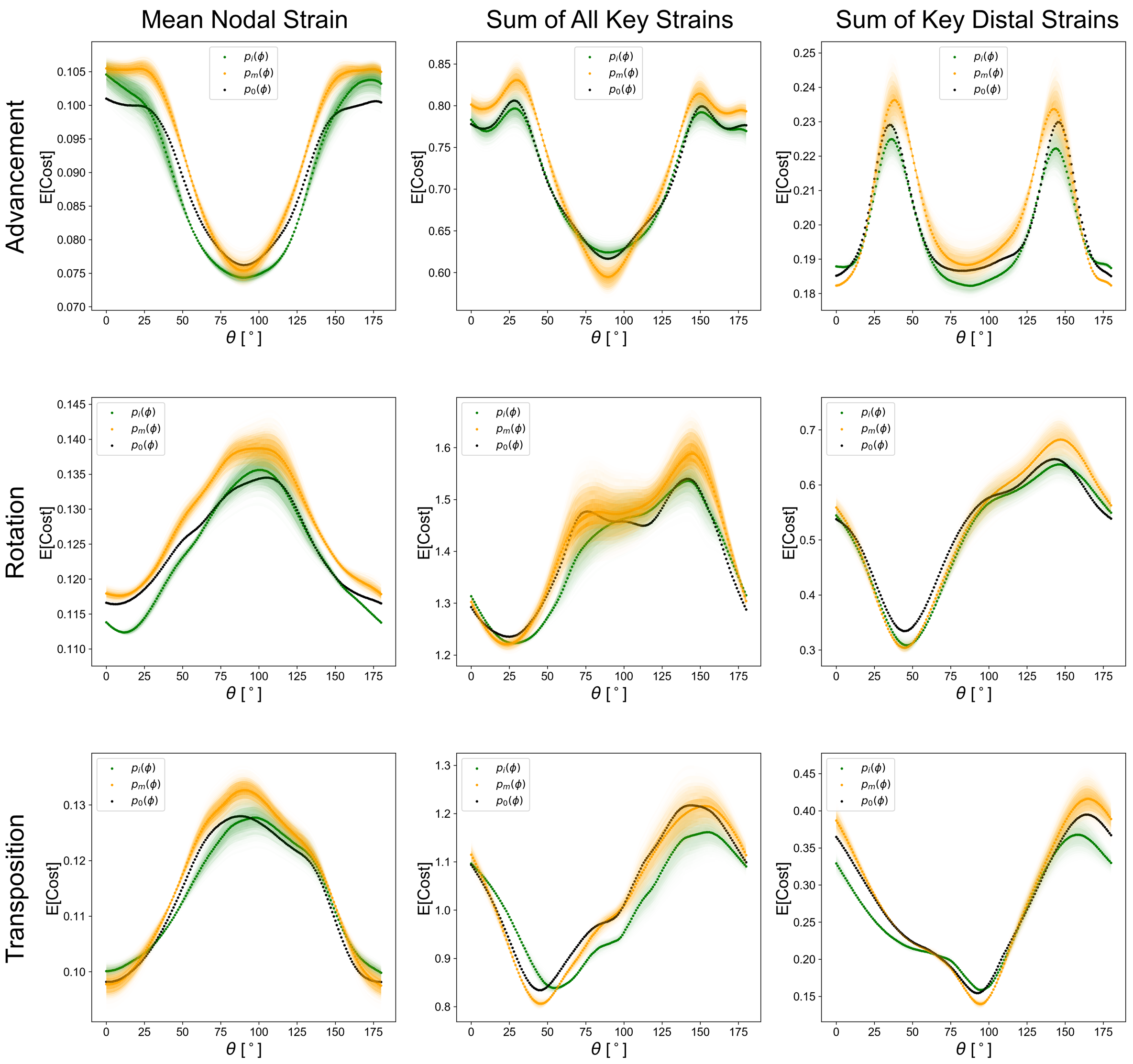}
\caption{Expected value of cost for 3 distributions of the material parameters $p(\phi)$: entire range, uniform distribution ($p_0(\phi)$); mean of range, normal distribution ($p_m(\phi)$); and 59 year old female, normal distribution ($p_i(\phi)$). The shaded regions around $p_m(\phi)$ and $p_i(\phi)$ indicate the cost distribution, while the solid line denotes the expectation of the cost. For the prior distribution $p_0(\phi)$ only the expected cost is shown.}
\label{fig09}
\end{center}
\end{figure*}

After evaluating the optimal design for the most general case in which no information is available for the material parameters $\phi$, we narrow our scope to two cases in which instead of using the prior $\phi_0(\phi)$, we assume that we have knowledge of individual parameter distributions for two cases. As described above, we solved the optimization problem described in \ref{eq_expected_cost} using $p_m(\phi)$, a normal distribution around the mean of the parameter ranges described in Table \ref{table01}, and also $p_i(\phi$ a narrow normal distribution around the parameters of a 59 year old female reported in \cite{tonge2013}. We plot in Fig. \ref{fig09} the expected value of the cost versus $\theta$ for these two distributions, $p_m(\phi)$ and $p_i(\phi)$ as well as the expected cost from the general study ($p_0(\phi)$). For the cases in which we sample from $p_m(\phi)$ and $p_i(\phi)$ we also show shaded regions around the expected cost to contrast the variation in the cost in these cases with respect to what is observed for $p_0(\phi)$ in Figures \ref{fig06}-\ref{fig08}. Immediately we note that the general trend of the curves for each of the individual cases is very similar to the general case in which the wide prior $p_0(\phi)$ was used. The values for $\theta$ that minimizes cost for each of these parameters fall within a small range. We also note that the area of uncertainty around the expected cost is much smaller than before, which is expected since now the parameters $\phi$ are sampled from narrower distributions.

\section*{Discussion}
The goal of the present manuscript was twofold. First, we sought to understand the impact of anisotropy on local reconstructive surgery flaps. Second, we aimed to develop computationally inexpensive surrogate models to quickly predict strain profiles for an arbitrary set of material parameters including anisotropy, enabling tasks such as optimization, uncertainty propagation, or model calibration needed in the clinical application. We focused on the three most common local reconstructive surgery flap designs - the advancement, rotation, and transposition flaps \cite{buchanan2016evidence,patel2011concepts}. 

The method relies first on thoroughly but efficiently exploring the input space of material parameters with a relatively small number of well-selected and semi-automated FE simulations, about 2000 per flap. The input space considered is five-dimensional, including anisotropy, and is based on a well-established hyperelastic model of collagenous tissues \cite{gasser2005,limbert2017}. The range of the parameters was based on the literature and spans up to three orders of magnitude for some parameters \cite{tonge2013,annaidh2012b,flynn2010simulating}. The output data from the detailed FE model is the high-dimensional strain field, on the order of tens of thousands of nodal values, which is not suitable for building the surrogates. We showed that PCA can reduce the dimensionality to a basis of approximately 30 features that capture more than 99$\%$ of the variance in the data. The GP surrogates trained on the reduced data can accurately predict the high-dimensional strain fields for any other input within the feasible range, as evidenced by the L2 norms of the error on the validation set. Unlike the original FE model, the surrogate is computationally inexpensive and enables fast prediction for arbitrary material properties, including anisotropy. We used the surrogates to perform a Sobol sensitivity analysis and determined that the fiber direction ($\theta$) is the most significant parameter to variations in the strain profiles. In view of the sensitivity analysis, we posed several optimization problems to identify the best orientation of the flaps with respect to anisotropy direction. 

\subsection*{Performance of GP surrogates}
Overall the GP surrogates performed well on the validation set, enabling fast prediction of the full strain fields for arbitrary choice of material properties. The advancement and transposition results were very accurate, with average errors of less than $2\%$, while the rotation results had larger errors and standardized residuals. There are numerous possible causes for this issue. One, the nodal strains for the rotation flap may vary more than those of the transposition and advancement flap. Thus, $N=2000$ training data sets may not have been sufficient to fully explore the outputs space for the rotation flap. Using more simulations for training may help overcome this issue. In addition, reducing the dimensionality of the output data with PCA does entail a loss of 1.0\% of the information, which could have been problematic for the rotation flap. Third, the suturing scheme for this flap leads to very intricate strain field, possibly singular, near the $Y$ junction at the base, which should be investigated further. A mesh sensitivity analysis was conducted for all flaps, but only for one set of  parameters; yet, it is possible that the mesh is not refined enough for other regions of the parameter space. Despite larger errors in the rotation flap with respect to the other two strategies, the average L2 norm of the relative error was still below 6.5$\%$ and this value was considered accurate enough for our purposes. 

In previous work, we also looked at the same flap designs - advancement, rotation, transposition- but we considered isotropic material properties \cite{lee2019}. In that work  we showed that a smaller training of approximately $N=1000$ simulations was enough to capture the response function. The material model used in \cite{lee2019} was the same as the one used here, yet, ignoring anisotropy, the input space consisted of only three parameters. The PCA step for the isotropic material revealed that 99.9$\%$ of the variance could be captured with less than 10 features. Here, adding two  parameters for anisotropy increased the dimension of the final PC basis, with roughly 30 features needed to describe $99\%$ of the variance. Compared to previous work, we increased the training data to $N=2000$, which was enough to produce very good results for the advancement and transposition surrogates, but only acceptable errors for the rotation case. While more FE simulations are likely to improve our predictions, we considered $N=2000$ to be a reasonable computational expense for the resulting accuracy, given that each simulation takes approximately three to five minutes on two Sky Lake CPUs at 2.60GHz. One of the main advantages of using GP surrogates is that they not only provide a prediction of the response function, but also of the uncertainty in that prediction. This information can be used, for example, to guide active learning strategies and run more simulations if needed, but only at points in the input space in which it would reduce the variance of the GP \cite{costabal2019}.

\subsection*{Optimal flap orientation}
One key result of this study is that the fiber direction ($\theta$) proved to be the most influential parameter on the variations in the strain profiles. This is significant because this parameter can be controlled clinically, unlike the other parameters. While a surgeon cannot control the direction of the skin's collagen fiber network, they can control the orientation of the flap with respect to the direction of anisotropy of skin. Clinically, the anisotropy direction for skin is described based on anatomy and commonly referred to as the direction of relaxed skin tension \cite{borges1984relaxed}. Thus, by orienting the flap in an optimal direction with respect to this direction of anisotropy, which we consider aligned with the underlying collagen network, the surgeon has some control over the resulting strain, ideally reducing complications in the healing process. While an initial optimization study was completed here, it would be useful to expand this study to gain a better understanding of the optimal value of $\theta$ for more realistic scenarios and refined cost functions. We posed three cost functions in light of clinical guidelines which suggest that surgery should minimize overall deformation, but especially deformation near the suture line and at distal ends of the flap \cite{barnhill1984study,leedy2005reconstruction,rohrer2005transposition,gurtner2011improving,buchanan2016evidence}. For the advancement flap, which has the most intuitive strain distribution, the result of our optimization aligns with the clinical recommendation of advancing perpendicular to the anisotropy direction. For the other two flaps, however, the strain distribution is more complex and the influence of the material properties is more noticeable. Our results suggest that for the rotation flap, angles in the approximate range $[10,40]^\circ$ are optimal. For the transposition case, we could not identify a consistent result based on the three different objective functions. In the transposition case, minimizing the average strain yields a different result compared to minimizing the sum of the strains near the suture region, and is also different from minimizing only the strains at the distal end of the flap. Therefore, more information is needed to refine the cost function that truly leads to the best clinical outcome. 

The initial optimization was done under the assumption that no information is available about the material properties of an individual. We then also explored the case in which some information is available. While having additional information reduced the uncertainty in the cost function, the results aligned with the more general case. Thus, we expect that our analysis with the most general distribution of material parameters should be useful to guide the flap design in the clinical setting, even when no information about skin material properties are available.

\subsection*{Limitations and ongoing work}

There are multiple limitations to this work that should be recognized. First, even though the surrogates are accurate over a broad input space covering the material response of skin, more input parameters need to be considered. For instance, we considered a single suturing scheme and a single set of boundary conditions. Future work includes incorporating changes in geometric parameters of the flaps, such as angle of transposition or angle of rotation for these two flap designs, or base to width ratio for the advancement flap. Suturing schemes can also be optimized and should thus be considered as inputs in metamodel creation, as well as the change in boundary conditions. 

Another limitation of the current work is the need for validation in the clinical setting. Numerical models of soft tissues are well-established \cite{mitchell2016real,buganza14a}. In particular regarding skin, the GOH hyperelastic material model has been deemed appropriate to capture the response of skin under tensile loading \cite{limbert2017,meador2020regional}. However, assumption of hyperelasticity is debatable for larger time scale for which the tissue may show viscoelastic behavior \cite{wahlsten2019compressibility}. The interaction between the skin and the underlying tissues is also ignored in this model, but could be important in a more realistic setting. The influence of pre-stress is also ignored in this study, but it could be an important factor \cite{rausch2013effect}. Finally, \textit{in vivo} measurem in the clinical setting  \cite{pittar2018scalp,pensalfini2018,muller2018novel}. Thus, while we are confident that the results shown here provide valuable insight into flap biomechanics and should be used to improve flap design, the models still need to be further validated with clinical data, which is the central task of our ongoing work. 

\section*{Conclusions}
In summary, we showed that a few detailed FE simulations, paired with reduced order modeling strategies, can be used to create inexpensive yet accurate GP surrogate models of the three most common flap designs. The surrogate models enable immediate prediction of strain profiles for arbitrary material properties including anisotropy, enabling tasks such as uncertainty propagation, model calibration, sensitivity analysis, and optimization. The direction of anisotropy with respect to the flap design is the single most important parameter that the surgeon can control to optimize a desired strain objective, even when the other material parameters are completely unknown.  

\section*{Supplement and code}
Code related to this publication is available at
\url{https://bitbucket.org/abuganzatepole/gp_anisotropy}

\section*{Acknowledgements}

This work was supported in part by the National Science Foundation under grant No.
1911346-CMMI and by the National Institute of Arthritis and Musculoskeletal and Skin Diseases of the National Institute of Health under award R01AR074525.

%







\end{document}